\documentclass[twocolumn]{article}
\usepackage{moreverb,url}
\usepackage[pdftex]{graphicx}
\usepackage{amsmath}
\usepackage{amssymb}
\usepackage{amsthm}
\usepackage[utf8]{inputenc}
\usepackage[T1]{fontenc}
\usepackage[colorlinks,bookmarksopen,bookmarksnumbered,citecolor=red,urlcolor=blue]{hyperref}
\usepackage{epstopdf}
\usepackage{subcaption}
\usepackage{smartref}

\newtheorem{thm}{Tétel}
\newtheorem*{def.}{Definition}
\newtheorem*{mj.}{Remark}
\newtheorem{lem.}[thm]{Lemma}
\newtheorem{prop}[thm]{Proposition}

\renewcommand\vec{\boldsymbol}

\newcommand{\m}[1]{\mathcal{#1}}
\newcommand{\norm}[1]{\left\lVert#1\right\rVert}
\newcommand{\abs}[1]{\left\lvert#1\right\rvert}

\begin{document}

\author{Tamás Baranyai}




\title{On the duality of space-trusses and plate structures of rigid plates and elastic edges}

\maketitle

\begin{abstract}
Dualities have been known to map space trusses and plate structures to each other since the 1980-s. Yet the computational similarity of the two has not been used to solve the unfamiliar plate structure with the methods of the well known truss. This paper gives a method to find the forces and displacements of a plate structure with rigid plates and elastic edges, using a dual truss. It is applicable for both statically determinate and indeterminate structures, subjected to both statical and kinematical loads. 
\end{abstract}

\section{Introduction}
Dualities as an articulated projective geometrical concept have emerged in the field of statics with Maxwell\cite{maxwell1870} and Cremona\cite{cremona1890graphical}, and were expanded at that time by Klein and Wieghardt\cite{klein1904onMaxwell}. While these methods were able to give one the forces of a (planar) truss, with the spread of algebraic methods they were almost forgotten in engineering circles. It was in the 1980-s when the renaissance of projective geometry in structural engineering started, discovering that the rigidity of a truss is a projective invariant \cite{crapo1982statics,wunderlich1982projective}. Investigation of the duality of engineering structures followed, both in 2D between plane trusses and grillages\cite{TARNAI1989duality,whiteley1991weavings,whiteley1989rigidity} and in 3D space between spatial trusses and plate (sheet) structures\cite{whiteley1987rigidity}. Parallel to this scientific renaissance, danish architect Wester\cite{wester1989design,wester1990geodesic} began researching and popularising plate structures for their efficiency and clarity ("pure plate action", in his words). 
While to this day very few homogeneous, cast concrete plate structures have been built, with the spread of automation into construction the use of smaller, prefabricated plates forming a spatial structure is getting more obvious and economical. This motivates reserchers to study, for instance, glass\cite{bagger2010plate} and wooden\cite{Krieg2015robotic} plate structures.
New ways are being tested to connect the prefabricated plates on-site. In case of glass this is usually done by glueing\cite{BLANDINI2007addhesive,SANTARSIERO2016laminated,ZANGENBERG201268}, while wooden elements easily allow the use of traditional finger joints and mechanical elements as well\cite{li2015segmental,stitic2018experimental}. In any case the on-site joints between the prefabricated elements are generally softer and they are generally responsible for the majority of the displacements occurring in the structure. This motivates the choice to model such structures with rigid plates connected by elastic joints.\\
The continuation of the research on projective geometry turned back to the path set by Maxwell, with its sight set on geometric representation of forces of spatial structures\cite{konstantatou2016reciprocal,mcrobie2016maxwell} and the extension of such diagrams to kinematics of the structures\cite{Mitchell2016partI,McRobie2016partII,McRobie2017elastographics}.
While La Magna et. al.\cite{lamagna2012nature} mention the possibility of using a dual truss structure to compute the edge forces of a plate structure, a method to  do this in the general case of a possibly statically indeterminate (hyperstatic) structure subjected to both statical and kinematical loads is hitherto undeveloped.

\section{Notation and modelling}
\subsection{Trusses and plate structures}
Here a quick review is given on the statical and compatibility equations of trusses and plate structures, presenting the similarity of the two structures and to provide the objects of comparison.
\subsubsection{The well known model of a truss}
is comprised of pin- or ball-joints and elastic bars connecting them. Loads are restricted to concentrated forces acting on the joints. The system of statical and compatibility equations take the form\cite{Livesley,roller_szabo} of:
\begin{align}
\begin{bmatrix}
0 & G'\\
G'^T & F'
\end{bmatrix}
\begin{bmatrix}
\epsilon\\
\phi'
\end{bmatrix}+
\begin{bmatrix}
\pi'\\
0
\end{bmatrix}=
\begin{bmatrix}
0\\
0
\end{bmatrix}.\label{eq:globt}
\end{align}
The upper block contains the equilibrium equations of the joints in global frames, while the lower block contains the compatibility equations of the bars in their local ($1$ dimensional) frames. In the case of a space-truss with $n$ joints and $l$ bars, $G'\in \mathbb{R}^{3n\times l}$ is the geometrical matrix while matrix $F'\in \mathbb{R}^{l\times l}$ is in fact diagonal and contains the stiffnesses of the bars on it's main diagonal. The vector $\epsilon \in \mathbb{R}^{3n}$  contains the displacements of the joints such that
\begin{align}
\epsilon=\begin{pmatrix}
\vec{\epsilon}_1\\
\vdots\\
\vec{\epsilon}_n
\end{pmatrix} \text{and} \ \vec{\epsilon}_i=\begin{pmatrix}
\epsilon_{i,1}\\
\epsilon_{i,2}\\
\epsilon_{i,3}
\end{pmatrix}.\label{globtr}
\end{align}

Here $\vec{\epsilon}_i$ is the 3 dimensional displacement of joint $i$. Vector $\pi'$ contains the loads of the vertices similarly. Vector
\begin{align}
\phi'=\begin{pmatrix}
\phi'_1\\
\vdots\\
\phi'_l
\end{pmatrix}
\end{align}

contains the bar forces, $\phi_i$ being the force in bar $i$.

\subsubsection{A common model of a plate structure}
\begin{figure*}
\begin{center}
\includegraphics[width=160mm]{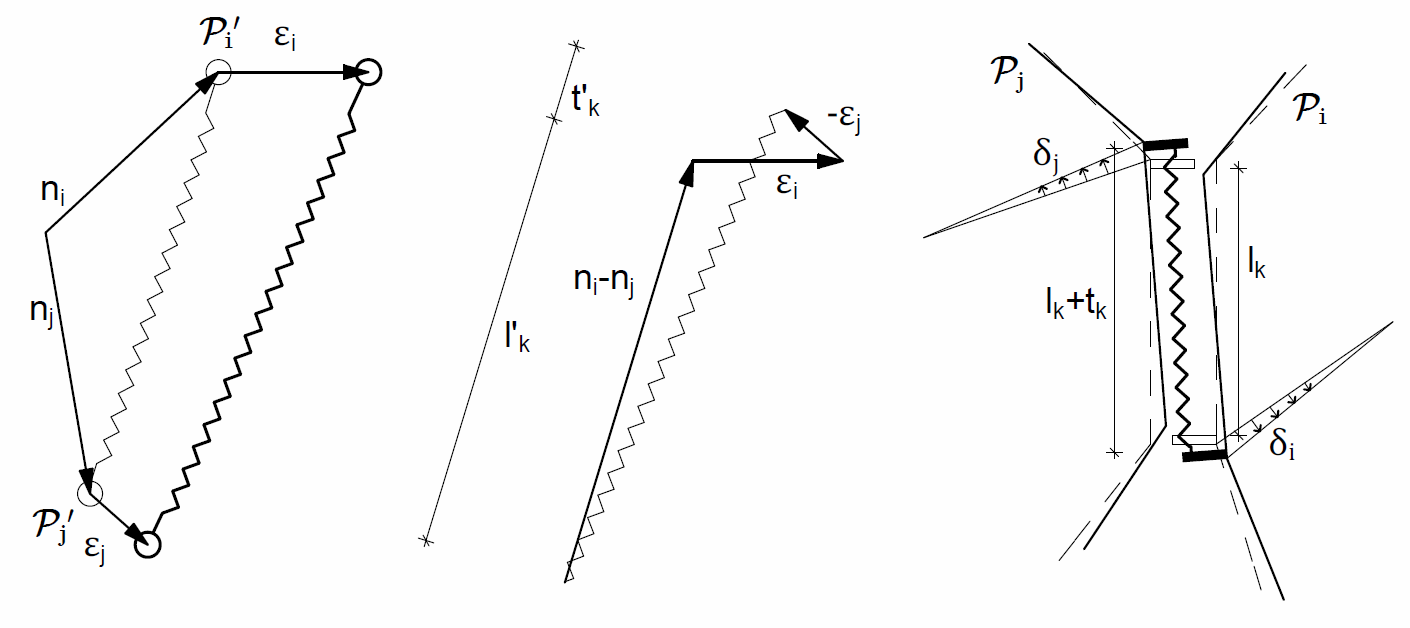}
\caption{Comparison of compatibility conditions. Left: displacements of joints of a truss. Middle: the change of barlength ($t'_k$) assuming small displacements. Right: rotational displacements of two plates of a plate structure.}\label{fig:kiselm}
\end{center}
\end{figure*}

is comprised of plates, subjected only to loads in their planes. In the following, the plates are considered to be rigid against forces acting in their planes, while their joins with other plates are considered to be elastic. The plates are considered to be so soft perpendicular to their planes that deformations in this direction happen without significant resistance. Hence only the displacements happening in the plane of the plates are relevant (cause forces) and the forces between two plates act such that they are in the common plane of both the plates they join. 
\begin{prop}\label{prop}
With the proper choice of equations and parameters the system of equilibrium and compatibility equations of a plate structure can be cast in the form:
\begin{align}
\begin{bmatrix}
0 & G\\
G^T & F
\end{bmatrix}
\begin{bmatrix}
\delta \\
\phi
\end{bmatrix}+
\begin{bmatrix}
\pi\\
0
\end{bmatrix}=
\begin{bmatrix}
0\\
0
\end{bmatrix}.\label{eq:globpl}
\end{align} 
\end{prop}
The top block contains the moment equations of the plates (in a coordinate system specified later) and the bottom block contains the compatibility equations of the edges in their local $1$ dimensional coordinate systems. Here also $G$ is the geometrical matrix, $F$ is the stiffness matrix of the elastic joins, $\phi$ contains the edge-forces with which the plates support each other, $\delta$ contains the small displacements (rotations) of the plates and $\pi$ contains the loads. While $G$ is more intuitive, $F$ might need to be clarified. If along edge $k$ two plates are joined with relative deformation $t_k$  (measured in the direction of the edge as translation) then the arising force $\phi_{k}$ acting on both plates is such that $t_k=F_{k,k}\abs{\phi_{k}}$ holds. The comparison of the compatibility conditions in case of a truss and the plate structure can be seen in Figure \ref{fig:kiselm}.\\

The similarity of \eqref{eq:globt} and \eqref{eq:globpl} suggests that analyzing a truss or a plate structure are mathematically equivalent. This enables one to determine the behaviour of a plate structure by investigating a properly chosen truss. This paper will show how this choice can be methodized with the canonical duality of projective geometry. The steps of the method are
\begin{enumerate}
\item Determine the geometry and stiffness of the dual truss
\item Determine the dual loads on the dual truss
\item Solve the dual truss
\item Transform the solution back to the plate structure
\end{enumerate}  

While projective geometry naturally gives duals of points, lines and planes, the transformation of forces, displacements and the stiffness matrix needs to be worked out. In this paper transformation matrices $\Delta$, $\Phi$ and $\Pi$ will be presented satisfying

\begin{align}
\delta&=\Delta \epsilon\\  \label{eq:mapsstart}
\phi&=\Phi \phi'\\
\pi&=\Pi \pi'\\
F&=\Phi^{-1} F' \Phi^{-1}. \label{eq:mapsend}
\end{align}

\subsection{Review of projective geometry and screw theory}
We will make use of concepts from projective geometry as well as screw theory. For the sake of clarity, they are provided here. Let us start by defining the following equivalence relation on a vector space $\mathbb{V}$:
\begin{align}
x\sim y \iff y\in\{\lambda x  \ \vert \ \lambda \in \mathbb{R}\setminus \{0\} \} \quad (x,y\in \mathbb{V}).\label{eq:equivalence}
\end{align}
Let us denote the arising equivalence class of $x$ with $x_{\sim}$. This notion enables us to assign (homogenous) coordinates to points, planes and lines of the projective space as follows.
 
\subsubsection{Homogenous coordinates of planes and points} can be attained by setting the dimension of the vectors in \eqref{eq:equivalence} to be $4$. Each equivalence class uniquely represents a point, and all points are represented this way. Although the projective space does not discriminate between its points, in human thought we often think of the three dimensional projective space $(PG(3))$ as a union of Euclidian space ($\mathbb{E}^3$) and the ideal plane at infinity (which is $PG(2)$ essentially). We will use the convention that if $\vec{p}\in \mathbb{E}^3$ we will use the representant of form  $(1,\vec{p})$, while in case of ideal points all representants are of form $(0,\lambda \vec{p})$. If no distinction is to be made we will use the form $(p_0,\vec{p})_{\sim}$.

Similarly, each plane is represented with a single equivalence class and each equivalence class represents a unique plane. Denoting the standard inner product with $\langle  \ , \ \rangle$, the incidence relations take the following form: point $\mathcal{P}$ with coordinates $p_{\sim}$ lies in plane $\mathcal{S}$ with coordinates $s_{\sim}$  if and only if $\langle p_{\sim},s_{\sim}\rangle=0$ holds. As it can be seen, different choice of representant from the same equivalence class does not affect this relation.

\subsubsection{Homogenous coordinates of lines}
are attained in a slightly different way. Let us have $\vec{l}=(l_1,l_2,l_3)$, $\vec{\bar{l}}=(l_4,l_5,l_6)$ such that the tuple $(\vec{l},\vec{\bar{l}})\in \mathbb{R}^6$, is not a 0 vector. The equivalence class $(\vec{l},\vec{\bar{l}})_{\sim}$ represents a line of $PG(3)$ if and only if 
\begin{align}
\langle l,\bar{l} \rangle=l_1 l_4+l_2 l_5+l_3 l_6 =0 \label{eq:plücker} 
\end{align}
holds, and all lines of $PG(3)$ are represented this way. This way of representation is called the Plücker-coordinates of a line and equation (\ref{eq:plücker}) the Plücker-identity.

%
%

 Here it is convenient to present a pair of additional identities, see Pottmann and Wallner \cite{pottmann2001computational} for more details:\\

Coordinates of the line passing through points $(p_0,\vec{p})_{\sim}$ and $(q_0,\vec{q})_{\sim}$  
are given by:
\begin{align}
(\vec{l},\vec{\bar{l}})_{\sim}=(p_0\vec{q}-q_0 \vec{p},\vec{p}\times \vec{q})_{\sim}. \label{eq:pont2e}
\end{align}
This returns $0$, iff the two points coincide. Similarly, the common line of planes $(u_0,\vec{u})_{\sim}$ and $(v_0,\vec{v})_{\sim}$ are given by 
\begin{align}
(\vec{l^*},\vec{\bar{l}^*})_{\sim}=(\vec{u}\times \vec{v},u_0\vec{v}-v_0 \vec{u})_{\sim}\label{eq:sik2e}
\end{align} 
which also gives $0$ iff the two planes are identical.

Observing equations (\ref{eq:pont2e}) and (\ref{eq:sik2e}) shows that any line passing through a given point arise from the linear combination of $3$ distinct lines passing through the point, and any line lying in a plane arises as a linear combination of $3$ distinct lines lying in the plane.

\subsubsection{Dualities}
are incidence preserving one-to one maps between points and planes of $PG(3)$. With the interchange of incidence relations (line joining points - line in which planes intersect) they induce a corresponding one to one mapping on lines. Dualities with period $2$ are called polarities and have often been used in graphical statics before\cite{maxwell1870,cremona1890graphical}. The effect of any duality on homogenous coordinates can be represented with an invertible matrix equivalence class $M_{\sim}$, such that the dual of point $(p_0,\vec{p})_{\sim}$ is  plane $(p_0,\vec{p})M_{\sim}$, and the dual of plane $(s_q,\vec{s})_\sim$ is point $(p_0,\vec{p})M^{-T}_{\sim}$. A corresponding linear map can be constructed acting on the Plücker-coordinates of lines.\\

The 'canonical' duality is the one represented with the $4$ dimensional identity matrix (its equivalence class), mapping point $(p_0,\vec{p})_{\sim}$ into plane $(p_0,\vec{p})_{\sim}$ and plane $(p_0,\vec{p})_{\sim}$ into point $(p_0,\vec{p})_{\sim}$. As it can be seen, it has period $2$ thus it is a polarity as well. (In  a more geometric approach it can be thought of as being represented by the purely complex unit sphere in the complex projective space.) The induced line-to line mapping in this case is given as  $(\vec{l},\vec{\bar{l}})_{\sim} \mapsto (\vec{\bar{l}},\vec{l})_{\sim}$.\\

So far we have homogenous coordinates, but we would like to do mechanics with metric quantities. In order to have some, let us introduce the notion of the oriented line segment (an idea is dating back to at least Klein\cite{clebsch1906mathematische}).
\begin{def.}
An oriented line segment $(\vec{l},\vec{\bar{l}})$ is a sliding-vector bound to line $(\vec{l},\vec{\bar{l}})_{\sim}$, with length $\norm{(\vec{l},\vec{\bar{l}})}$ and a direction, which are given by:
\begin{itemize}
\item if $\norm{\vec{l}}>0$, then $\norm{(\vec{l},\vec{\bar{l}})}=\norm{\vec{l}}$ and the direction is given by $\vec{l}$ (finite line).
\item if $\norm{\vec{l}}=0$, then $\norm{(\vec{l},\vec{\bar{l}})}=\norm{\vec{\bar{l}}}$ and the direction is given by $\vec{\bar{l}}$ (ideal line). 
\end{itemize}
\end{def.}

We can now give the effect of the duality on this metric quantity similarly to the line, as:
\begin{align}
(\vec{l},\vec{\bar{l}}) \mapsto (\vec{\bar{l}},\vec{l}).
\end{align}

\subsubsection{Screw theory}
was originally proposed by Sir Robert Ball\cite{ball1998treatise}, providing a connection between Plücker coordinates of lines and force/velocity/displacement systems. Here a quick excerpt is presented, the reader may find the modern (engineering) interpretation in Davidson and Hunt\cite{davidson2004robots} or Gallardo-Alvarado\cite{gallardo2016kinematic}. For the mathematical minded, there is Pottmann and Wallner\cite{pottmann2001computational} and even Felix Klein\cite{clebsch1906mathematische}.\\

The effect of system of forces and moments in $\mathbb{E}^3$ can be given (after a choice of coordinate system) as the vector couple $(\vec{R},\vec{T})$, where $\vec{R}$ is a force acting at the chosen origin and $\vec{T}$ is a moment both containing the original moments and accounting for the translation of the elements of the force system into the chosen origin. Generally this vector couple is called a dyname or wrench. In the special case of $\langle\vec{R},\vec{T}\rangle=0$ (the Plücker identity, see \eqref{eq:plücker}) the dyname can be reduced into a single force (if $\vec{R}\neq \vec{0}$) or a single moment (if $\vec{R}=\vec{0}$). In the first case  $(\vec{R},\vec{T})$ is a representant from the $(\vec{R},\vec{T})_{\sim}$ equivalence class, corresponding to a regular oriented line segment in $PG(3)$, which is the line of action of the resulting force. In the second case it is representing a line segment lying in the ideal plane (see \eqref{eq:pont2e}). In the following we will consider only dynames satisfying \eqref{eq:plücker} and think of moments as forces lying in the ideal plane, acting along the corresponding ideal lines.\\

By using oriented line segments, a force $\vec{F}$ acting along line $(\vec{l},\vec{\bar{l}})_{\sim}$ can be represented as
\begin{align}
\frac{\norm{\vec{F}}}{\norm{(\vec{l},\vec{\bar{l}})}} (\vec{l},\vec{\bar{l}}),
\end{align}
assuming the force and the line segment point to the same direction.

Although in screw theory mostly forces and instantaneous velocities are represented as screws, small displacements can be handled this way as well, since they behave similarly to instantaneous velocities. We only need the formulation with the oriented line segment, as follows:
One can describe a small rotation with angle $\alpha$ around axis $(\vec{l},\vec{\bar{l}})_{\sim}$ as
\begin{align}
\frac{\alpha}{\norm{(\vec{l},\vec{\bar{l}})}} (\vec{l},\vec{\bar{l}}).
\end{align}
The sign of $\alpha$ is according to the handedness of the coordinate system.\\

The following operation will be used to describe the relation of forces and the displacements they cause:

\begin{align}
(\vec{l},\vec{\bar{l}})\circ (\vec{m},\vec{\bar{m}}):=\langle\vec{l},\vec{\bar{m}}\rangle+\langle\vec{\bar{l}},\vec{m}\rangle.
\end{align}

From the mathematical point of view, this is an indefinite inner product, satisfying linearity, commutativity but not positive definiteness. Also, (different) lines $(\vec{l},\vec{\bar{l}})_\sim$ and $ (\vec{m},\vec{\bar{m}})_\sim$ intersect if and only if $(\vec{l},\vec{\bar{l}})_\sim \circ (\vec{m},\vec{\bar{m}})_\sim=0$. From the physical point of view this operation can represent many things. The relevant ones are:

\begin{itemize}
\item The quantity $\frac{\norm{\vec{F}}}{\norm{(\vec{l},\vec{\bar{l}})}} \frac{1}{\norm{(\vec{m},\vec{\bar{m}})}} (\vec{l},\vec{\bar{l}})\circ(\vec{m},\vec{\bar{m}})$ is the moment of force $\frac{\norm{\vec{F}}}{\norm{(\vec{l},\vec{\bar{l}})}} (\vec{l},\vec{\bar{l}})$ to the axis $(\vec{m},\vec{\bar{m}})_{\sim}$
\item Given a rigid body rotated with  $\frac{\alpha}{\norm{(\vec{m},\vec{\bar{m}})}} (\vec{m},\vec{\bar{m}})$, the quantity $\frac{\alpha}{\norm{(\vec{m},\vec{\bar{m}})}} \frac{1}{\norm{(\vec{l},\vec{\bar{l}})}} (\vec{m},\vec{\bar{m}})\circ(\vec{l},\vec{\bar{l}})$ is the $(\vec{l},\vec{\bar{l}})$ directional component of the displacement of all points of the body lying on $(\vec{l},\vec{\bar{l}})_{\sim}$
\end{itemize}

Note how the canonical duality satisfies the following: Given two line segments $(\vec{l},\vec{\bar{l}})$ and $(\vec{m},\vec{\bar{m}})$, as well as their duals $(\vec{l},\vec{\bar{l}})'$ and $(\vec{m},\vec{\bar{m}})'$, we have:
\begin{align}
(\vec{l},\vec{\bar{l}})\circ (\vec{m},\vec{\bar{m}})=(\vec{l},\vec{\bar{l}})'\circ (\vec{m},\vec{\bar{m}})'\label{eq:circinv}
\end{align}
which will greatly simplify our analysis.

\subsection{Bases and dual bases}
Due to the linear nature of the problem, the effect of the duality can be described with its effect on appropriately chosen bases. First a plate in plane $\mathcal{P}$ and joint at dual point $\mathcal{P}'$ will be considered. Let us have an orthonormal base in the finite part of $PG(3)$, given by $\vec{b_1}$, $\vec{b_2}$ and $\vec{b_3}$, such that the origin is neither $\mathcal{P}'$ nor lying in $\mathcal{P}$. With this restriction both point and plane can be represented with the same vector: $(1,\vec{n})$. We will also consider the additional condition of $\vec{b}_1\parallel \vec{n}$, and we will take note of what it changes. Now two pairs of bases can be created, bound to $\mathcal{P}$ and $\mathcal{P}'$ respectively.

\begin{figure*}
\centering
\includegraphics[width=160mm]{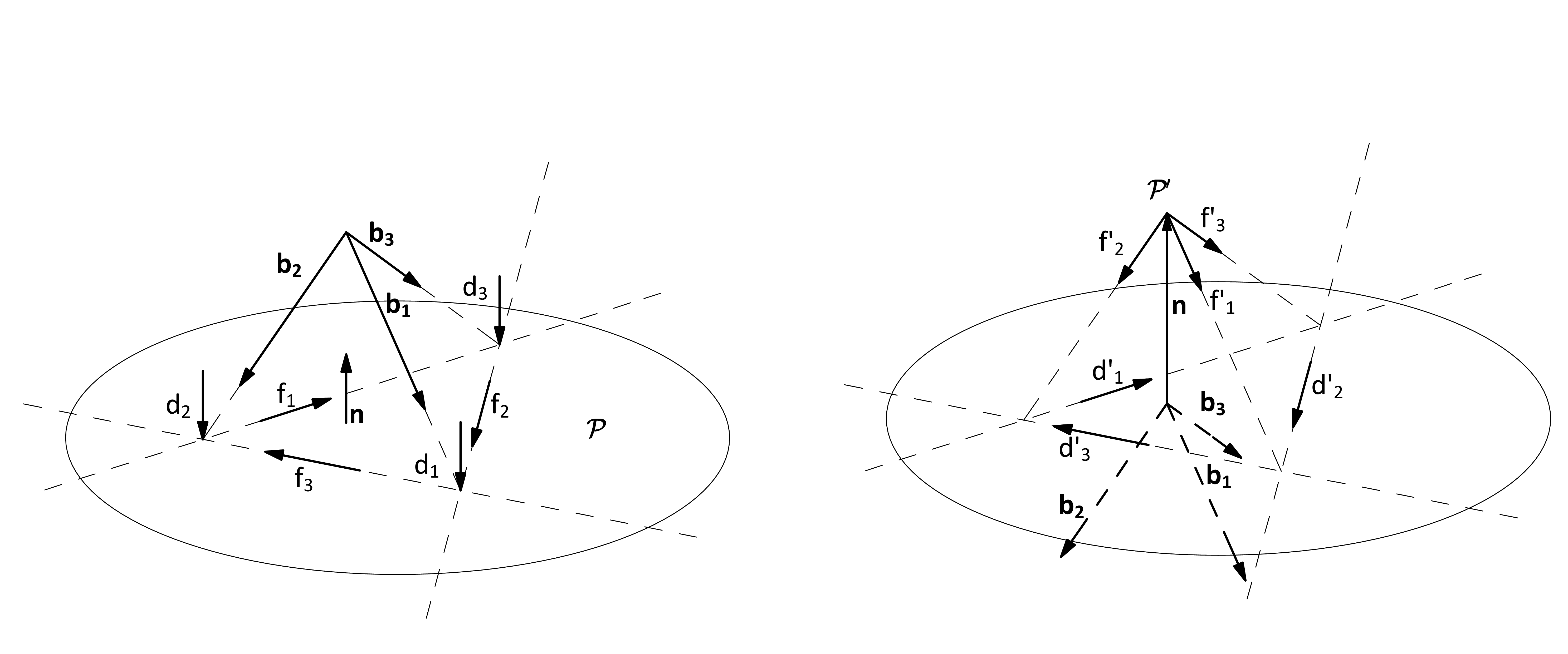}
\caption{left:Bases associated with plane $(1,\vec{n})$. Elements of $\{f_j\}$ are lying in it while elements of $\{d_j\}$ are perpendicular to it. right:Dual bases, associated with point $(1,\vec{n})$. Elements of $\{f'_j\}$ are passing through it, while elements of $\{d'_j\}$ lie on a plane passing through the origin of $\{\vec{b_j}\}$.}\label{fig:bazisegyben}
\end{figure*}

\begin{figure*}
\centering
\includegraphics[width=160mm]{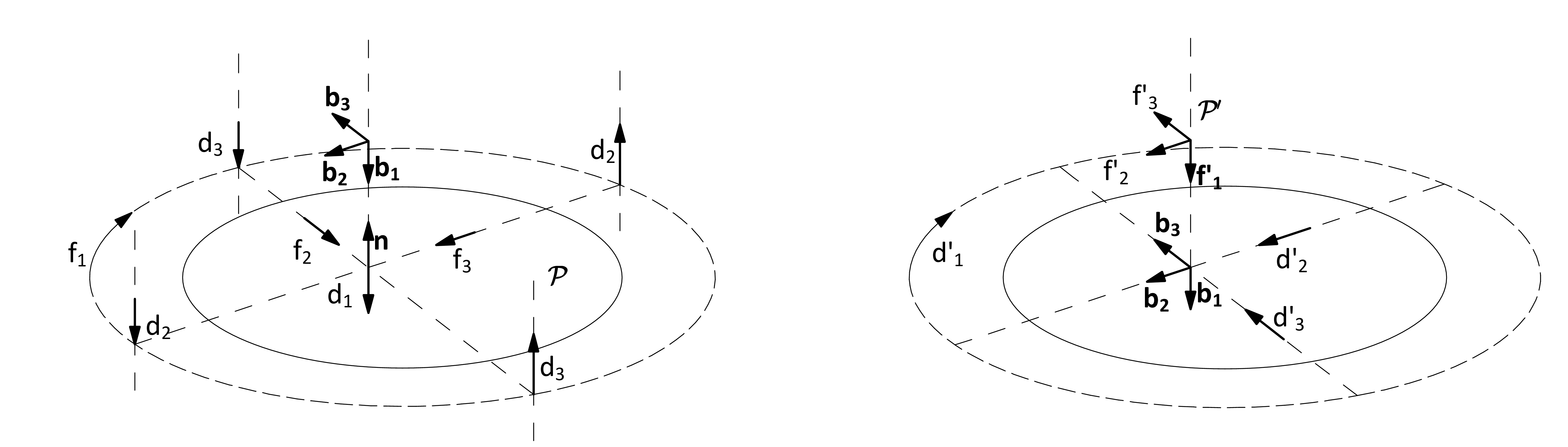}
\caption{Bases an dual bases if $\vec{b}_1\parallel \vec{n}$ is satisfied. Some line segments are drawn twice, their appearently opposite diretion is attributed to the fact that all lines are loops in $PG(3)$. Left: bases associated with plane $(1,\vec{n})$. Line segment $f_1$ is along the ideal line of the plane, acounting for the moments in the plane. Line segments $d_2$ and $d_3$ also lie in infinity, representing pure translations of the plane. Right:dual bases, associated with point $(1,\vec{n})$. Line segment $d'_1$ lies along an ideal line.}\label{fig:bazisegyben_spec}
\end{figure*}

The first one will be spanned by $d_1$, $d_2$ and $d_3$, where
\begin{align}
d_j:=(\langle \vec{n},\vec{b_j}\rangle \vec{n},\vec{n}\times\vec{b_j}) \quad j=1\dots 3.
\end{align}
All line segments perpendicular to $\m{P}$ can be given as a linear combination of these, since they all pass through $(0,\vec{n})$ and are linearly independent. Thus any rotation (or translation) of $\m{P}$ can be given as
\begin{align}
\sum_{j=1 \dots 3} \frac{\mu_{j}}{\norm{d_{j}}} d_{j} \quad \vert \ \mu_j \in \mathbb{R}.
\end{align} 
If the additional condition of $\vec{b}_1\parallel \vec{n}$ is satisfied, than $d_2$ and $d_3$ intersect $\m{P}$ in ideal points, thus pure translations of the plate in the plane are represented with linear combinations where $\mu_1=0$. 

The next base will be spanned by $f_1$, $f_2$ and $f_3$, where
\begin{align}
f_j:=(\vec{n}\times\vec{b_j},\vec{b_j}) \quad j=1\dots 3.
\end{align}
Any oriented line segment lying in plane $\m{P}$ can be given as a linear combination of these (see \eqref{eq:sik2e}). Thus any force acting in $\m{P}$ can be given as
\begin{align}
\sum_{j=1 \dots 3} \frac{\mu_{j}}{\norm{f_{j}}} f_{j} \quad \vert \ \mu_j \in \mathbb{R}.
\end{align}
The satisfaction of the additional condition of $\vec{b}_1\parallel \vec{n}$ means $f_1$ is the ideal line of the plane and a moment acting in the plane is given by $\mu_2=\mu_3=0$, while $\mu_1=0$ means the force is passing through the point where $d_1$ intersects $\m{P}$.\\

The corresponding dual base for forces passing through $\m{P}'$ will be $\{f'_j\} $ where
\begin{align}
f'_j:=(\vec{b_j},\vec{n}\times\vec{b_j}) \quad j=1\dots 3.
\end{align}
It's easy to see (from \eqref{eq:pont2e}) that any force passing through $\m{P'}$ can be given as a linear combination of these. The dual base for the displacements of the joint at $\m{P'}$ will be $\{d'_j\} $ where
\begin{align}
d'_j:=(\vec{n}\times\vec{b_j},\langle \vec{n},\vec{b_j}\rangle \vec{n}) \quad j=1\dots 3.
\end{align}
They all lie in a plane passing through the origin, perpendicular to $\vec{n}$. Any displacement of $\m{P}$ can be expressed with the help of these, such that:

\begin{align}
\sum_{j=1 \dots 3} \frac{\mu_{j}}{\norm{d'_j}} d'_j  \quad \vert \mu_j \in \mathbb{R} \label{eq:dispdesc}
\end{align}

To see that this indeed spans $\mathbb{E}^3$ and how the displacements scale one can express the translation caused at $\m{P}'$ as
\begin{align}
\sum_{k=1 \dots 3}\vec{b}_k\sum_{j=1 \dots 3}\frac{\mu_{j}}{\norm{d'_{j}}\norm{f'_{k}}} d'_{j}\circ f'_{k}.
\end{align}
Where
\begin{align}
\begin{split}
d'_{j}\circ f'_{k} =&\langle(\vec{n}\times\vec{b}_j),(\vec{n}\times\vec{b}_k)\rangle +\\
&+ \langle\vec{n},\vec{b}_j\rangle \langle \vec{n},\vec{b}_k\rangle\end{split} \label{eq:algprrof}\\
\begin{split}
=&\langle\vec{n},\vec{n}\rangle \langle \vec{b}_j,\vec{b}_k\rangle-\\
&-\langle\vec{n},\vec{b}_j\rangle \langle \vec{n},\vec{b}_k\rangle+\\
&+\langle\vec{n},\vec{b}_j\rangle \langle \vec{n},\vec{b}_k\rangle\end{split}\\
=&\langle\vec{n},\vec{n}\rangle \langle \vec{b}_j,\vec{b}_k\rangle. \label{eq:algprrofend}
\end{align}
This, the orthonormality of $\{\vec{b}_j\}$ and the fact that $\forall \ j \ \norm{f'_j}=1$ shows that the translation of $\m{P}$ from the rotation described in \eqref{eq:dispdesc} is 
\begin{align}
\norm{\vec{n}}^2\sum_{j=1 \dots 3} \frac{\mu_{j}}{\norm{d'_{j}}} \vec{b}_j.\label{eq:vetuletkapcsolat}
\end{align}
With this and \eqref{eq:circinv} we have also discovered which base vector (line segment) intersects which one. Here the satisfaction of the additional condition of $\vec{b}_1\parallel \vec{n}$ implies that $d'_1$ lies at infinity. A drawing of the bases is presented in Figure \ref{fig:bazisegyben} in the general case and in Figure \ref{fig:bazisegyben_spec} if $\vec{b}_1\parallel \vec{n}$ is satisfied.
\begin{mj.}
The restriction of the origin of $\{b_j\}$ not being a vertex of the truss or in a plane of a plate is a necessary condition of having two finite dual structures.
\end{mj.}

\subsection{Extension to multiple plates and joints}
Trusses and plate structures consist of multiple plates/joints and edges/bars. The vector representing plane $\m{P}_i$ and point $\m{P}'_i$ will be denoted with $(1,\vec{n}_i)$. Consequently there will be multiple bases attached to points and planes, for instance the $j$-th element of the base for forces attached to $\m{P}_i$ will be denoted with $f_{i,j}$. The extension of the additional condition on base directions takes the form of $\vec{b}_{i,1} \parallel \vec{n}_i$ for all $i$. Moreover, a line segment along edge $k$ and dual bar $k$ will be denoted with $e_k$ and $e'_k$ respectively.

Now the validity of Proposition \ref{prop} can be shown. If edge force $\phi_k$ acts along edge $e_k$, its moment to axis $d_{i,j}$ is given by
\begin{align}
\phi_k \frac{e_k \circ d_{i,j}}{\norm{e_k}\norm{d_{i,j}}}.
\end{align}
Similarly, the $e_k$ directional component from the small rotation $\delta_{i,j}$ around axis $d_{i,j}$ is
\begin{align}
\delta_{i,j} \frac{e_k \circ d_{i,j}}{\norm{e_k}\norm{d_{i,j}}}.
\end{align}
It is apparent that if we want to use the same matrix (although transposed) in both the equilibrium and the compatibility equations, the required equilibrium equations are the moment equations around axes $d_{i,j}$. By introducing the notation $r=3(i-1)+j$, the elements of $G$ can be given as
\begin{align}
G_{r,k}=
\begin{cases} 
      \frac{e_k \circ d_{i,j}}{\norm{e_k}\norm{d_{i,j}}} & \text{if }e_k \text{ is an edge of the plate } \m{P}_i \\
      0 & \text{otherwise}. 
\end{cases}
\end{align}

In the following we will describe how statics and kinematics of a plate structure translate to statics and kinematics of the dual truss.

\section{Finding the transformation matrices} 
\subsection{Transformation of forces}
Consider plate $\m{P}_i$ acted upon by force 
\begin{align}
\sum_{j=1 \dots 3} \frac{\pi_{i,j}}{\norm{f_{i,j}}} f_{i,j}
\end{align} and supported along its $e_k$ edges with $\phi_k$ edge forces ($k \in \m{I}$ some index-set). The equilibrium of forces is captured with the equation:
\begin{align}
\sum_{j=1 \dots 3} \frac{\pi_{i,j}}{\norm{f_{i,j}}} f_{i,j}+\sum_{k \in \m{I}} \frac{\phi_{k}}{\norm{e_{k}}} e_{k}=0.\label{eq:stat}
\end{align}
In case of the dual vertex at point $\m{P}'$ the equilibrium of the dual forces and the dual load gives the equation:
\begin{align}
\sum_{j=1 \dots 3} \frac{\pi'_{i,j}}{\norm{f'_{i,j}}} f'_{i,j}+\sum_{k \in \m{I}} \frac{\phi'_{k}}{\norm{e'_{k}}} e'_{k}=0.\label{eq:dstat}
\end{align}
Both \eqref{eq:stat} and \eqref{eq:dstat} are linear combinations where the corresponding vectors ($f_{i,j}$ to $f'_{i,j}$ and $e_{k}$ to $e'_{k}$) differ only in a common permutation. This implies that given the additional constraints
\begin{align}
\frac{\pi_{i,j}}{\norm{f_{i,j}}}=\frac{\pi'_{i,j}}{\norm{f'_{i,j}}} \quad \text{and} \quad \frac{\phi_{k}}{\norm{e_{k}}}=\frac{\phi'_{k}}{\norm{e'_{k}}} \label{eq:statconst}
\end{align}
the two equilibrium equations are equivalent. Since this is precisely what we want, we determine $\Pi$ and $\Phi$ such that \eqref{eq:statconst} is satisfied. Both $\Phi$ and $\Pi$ are diagonal, the $k,k$-th element of $\Phi$ is 
\begin{align}
\Phi_{k,k}=\frac{\norm{e_k}}{\norm{e'_k}},
\end{align}
while $\Pi$ is more conveniently given in block-diagonal form as
\begin{align}
\Pi=
\begin{bmatrix}
\Pi_1 & &  &  &  \\
 & \ddots &  &  \\
 &  & \Pi_i &  \\
 &  &  & \ddots \\
 &  &  & & \Pi_n
\end{bmatrix}
\end{align}
where
\begin{align}
\Pi_i=
\begin{bmatrix}
\norm{f_{i,1}} & & \\
 & \norm{f_{i,2}} &  \\
 &  &  \norm{f_{i,3}}
\end{bmatrix}
\end{align}
since $\norm{f'_{i,j}}=1$ for all $i$ and $j$.\\
Solving equation (\ref{eq:globt}) tells us the magnitude of bar forces, and whether they represent tension or compression. In order to use the dual structure effectively, one must see how the sign convention appears in the plate structure.
If bar $k$ joining vertices $\vec{n}_i$ and $\vec{n}_j$ is under tension, the effect of its force on vertex $i$ can be given as 
\begin{align}
\frac{\phi'_{k}}{\norm{\vec{n}_j-\vec{n}_i}}(\vec{n}_j-\vec{n}_i,\vec{n}_i\times\vec{n}_i)
\end{align}  
such that $\phi'_{k}>0$. This means the dual force is of form 
\begin{align}
\frac{\phi_{k}}{\norm{\vec{n}_i\times\vec{n}_j}}(\vec{n}_i\times\vec{n}_j,\vec{n}_j-\vec{n}_i)
\end{align}  
such that $\phi_{k}>0$. This segment is pointing towards $\vec{n}_i\times\vec{n}_j$. We may observe then that if bar $k$ is under tension, then the dual edge force acting on plate $i$ is such that $\vec{n_i}$, $\vec{n_j}$ and the edge force $\phi_k(\vec{n}_i\times\vec{n}_j)$ are ordered according to the handedness of the coordinate system. Interchanging $\vec{n_i}$ and $\vec{n_j}$ and remembering the anti-commutativity of the cross product shows, this satisfies Newton's third law of forces and reactions. While $\vec{n}_i\neq\vec{n}_j$ is a trivial requirement, $\vec{n}_i \parallel \vec{n}_j \iff \norm{\vec{n}_i \times \vec{n}_j}=0$ would cause problems. In this case however, the intersection of the planes would lie in the ideal plane. Since we consider only finite structures, this never happens and the provided explanation is general enough. 

\subsection{Transformation of displacements}
The displacement $d_i$ of $\m{P}_i$ can be given as 
\begin{align}
d_i=\sum_{j=1 \dots 3} \frac{\delta_{i,j}}{\norm{d_{i,j}}} d_{i,j},\label{eq:daw_disp1}
\end{align} 
while the dual displacement $d'_i$ is
\begin{align}
d'_i=\sum_{j=1 \dots 3} \frac{\delta'_{i,j}}{\norm{d'_{i,j}}} d'_{i,j}.\label{eq:daw_disp2}
\end{align}
Similarly to the case of the forces, these are linear combinations in which the vectors differ only in a common permutation. Furthermore, the method we want to create is good if it maps unit displacement to unit displacement as: 

\begin{align}
\frac{1}{\norm{d_{i,j}}} \mapsto \frac{1}{\norm{d'_{i,j}}}
\end{align}
implying
\begin{align}
\frac{\delta_{i,j}}{\norm{d_{i,j}}} = \frac{\delta'_{i,j}}{\norm{d'_{i,j}}}.
\end{align}

However, what we usually prefer in case of a truss is the translational displacement of $\m{P}_i'$,  in the form of $\sum_{j=1 \dots 3} \epsilon_{i,j} \vec{b}_{i,j}$. We can use \eqref{eq:vetuletkapcsolat} to get precisely that and have
\begin{align}
\delta_{i,j}=\frac{\norm{d_{i,j}}}{\norm{\vec{n}_i}^2}\epsilon_{i,j}.\label{eq:delta-epsilon}
\end{align}

From this we see that the matrix $\Delta$ can also be given in block-diagonal form, as
\begin{align}
\Delta=
\begin{bmatrix}
\Delta_1 & &  &  &  \\
 & \ddots &  &  \\
 &  & \Delta_i &  \\
 &  &  & \ddots \\
 &  &  & & \Delta_n
\end{bmatrix}
\end{align}
where
\begin{align}
\Delta_i=\frac{1}{\norm{\vec{n}_i}^2}
\begin{bmatrix}
\norm{d_{i,1}} & & \\
 & \norm{d_{i,2}}  &  \\
 &  &  \norm{d_{i,3}}
\end{bmatrix}.
\end{align}

While we see that $\Phi$ is determined only by the geometry of the structure and the center of the duality (the origin of the coordinate system in this case), $\Pi$ and $\Delta$ are sensitive to directions of $\vec{b}_{i,j}$. More precisely, if the additional condition of $\vec{b}_{i,1} \parallel \vec{n}_i$ holds, then $\Pi^{-1}=\Delta$ is true.

\subsection{Transformation of the system of equations}
Now we can examine how and when the systems of equations can be transformed into each other. In case of the equilibrium equations, we can write

\begin{align}
G\phi+\pi=0\\
G\Phi\phi'+\Pi \pi'=0\\
\Pi^{-1}G\Phi \phi' + \pi'=0
\end{align}
implying 
\begin{align}
G'= \Pi^{-1} G \Phi.\label{eq:cont1}
\end{align}
In case of the compatibility equations, this is
\begin{align}
G^T \delta + F \phi=0\\
G^T \Delta \epsilon+F \Phi \phi'=0\\
\Phi G^T \Delta\epsilon+\Phi F \Phi \phi'=0,
\end{align}
which would imply 
\begin{align}
G'=\Delta G \Phi\label{eq:cont2} \\
F'=\Phi F \Phi.
\end{align}

Two things are noteworthy. Firstly, the connection between $F$ and $F'$ is determined only by the geometry of the structure and the center of the polarity, not the choice of bases. Secondly, if the bases are chosen such that $\vec{b}_{i,1} \parallel \vec{n}_i$ (or $\vec{b}_{i,j} \parallel \vec{n}_i$ for all $i$ and any $j$) is satisfied, then there is no contradiction between \eqref{eq:cont1} and \eqref{eq:cont2} and the whole system of equations can be transformed.

This condition however is not necessary to use this method. In fact it might be useful to use bases given by assuming $\vec{b}_{i,j}=\vec{b}_{k,j} \quad \forall i,k\in \{1 \dots n \}$, resulting in the frame at each joint having the same orientation. The matrices $\Pi,\Delta$ and $\Phi$ can be constructed as given, and they can be used to transform the loads and the stiffness matrix. The dual truss can be solved with whatever equations one likes, and the resulting forces and displacements can be transformed back.

\section{A numerical example}
\begin{figure}[h]
\begin{center}
\includegraphics[width=80mm]{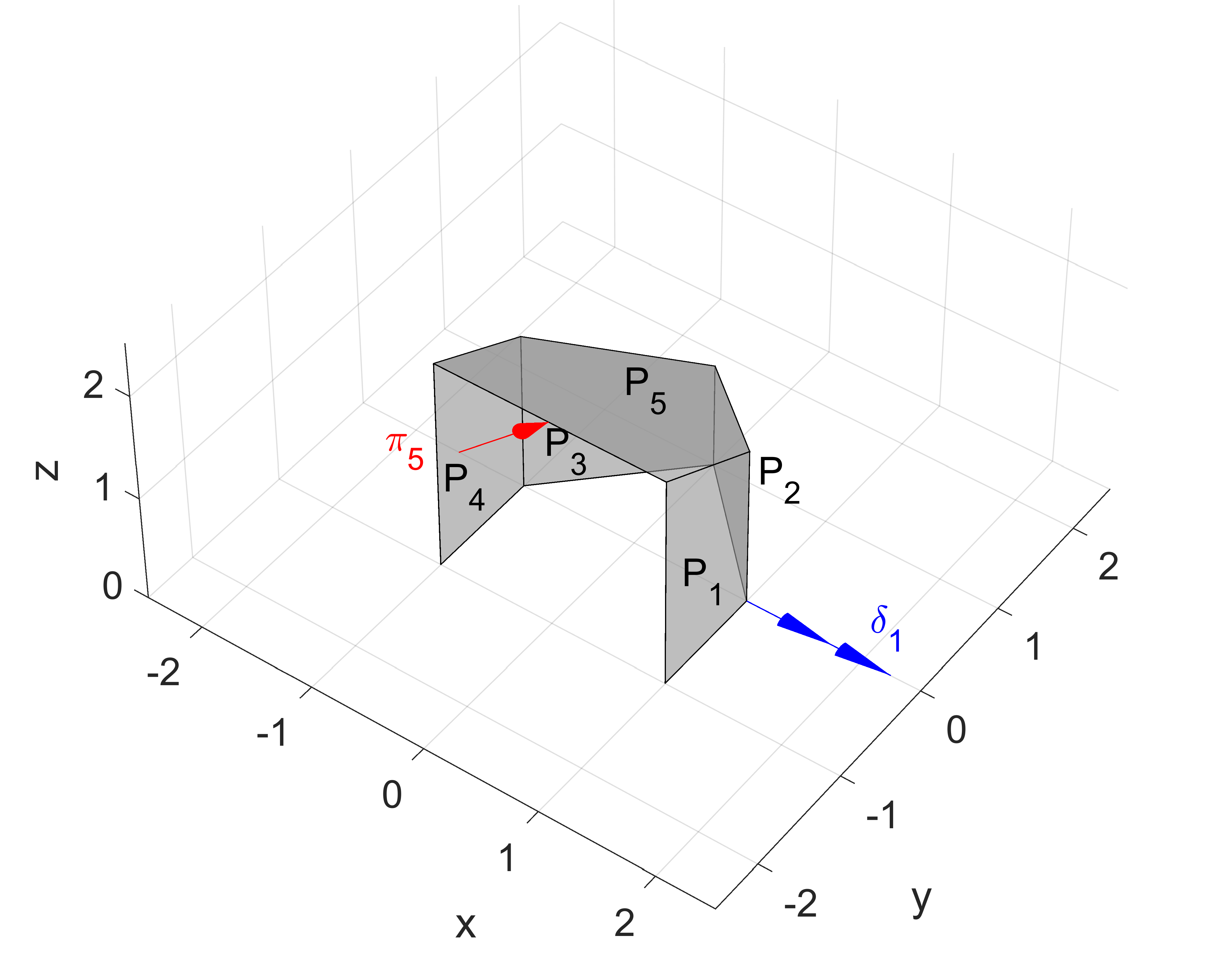}
\caption{Plate structure arising from 5 plates (the vertical ones are fixed, supporting ones) and 4 internal edges}\label{fig:lappelda}
\end{center}
\end{figure}

Consider the plate structure seen in Figure \ref{fig:lappelda}. The vertical plates are fixed in their planes supporting plate $5$ in a statically indeterminate way. The force $\pi_5$ is of magnitude $1$ kN, and there is an additional support displacement: plate $1$ is rotated around the $x$ axis with $1/10$ degrees ($0.0017$ radians).

The geometry of the plates is given by vectors:
\begin{align}
(1,\vec{n}_1)&=(1,-1,0,0)\\
(1,\vec{n}_2)&=(1,-1,-1,0)\\
(1,\vec{n}_3)&=(1,1,-1,0)\\
(1,\vec{n}_4)&=(1,1,0,0)\\
(1,\vec{n}_5)&=(1,0,-0.2,-0.4).
\end{align}

The plates are assumed to be glued together with polyurethane glue in $1$ mm thickness and in $2$ cm width. By assuming the glue has shear modulus of $1$ N/mm$^2$ and after working out the edge lengths the stiffness matrix takes the form of
\begin{align}
F=
\begin{bmatrix}
    0.4472  &       0 &        0 &        0\\
         0  & 0.3333  &       0  &       0\\
         0  &       0 &   0.3333 &        0\\
         0  &       0 &        0 &   0.4472\\
\end{bmatrix}10^{-7}\frac{\text{m}}{\text{kN}}.
\end{align}
The internal edges will be numbered such that edge $k$ joins plate $k$ with plate $5$, this way the edges can be represented with
\begin{align}
e_{1}= (0 ,  -0.4 ,   0.2  , 1,  -0.2  , -0.4)\\
e_{2}=(0.4 ,  -0.4  ,  0.2  ,  1  ,  0.8,-0.4)\\
e_{3}=(0.4, 0.4, -0.2,  -1,  0.8,  -0.4)\\
e_{4}=( 0,    0.4, -0.2, -1, -0.2, -0.4),
\end{align}
and from this the matrix relating bar and edge forces is:
\begin{align}
\Phi=
\begin{bmatrix}
    0.4082  &       0    &     0 &        0\\
         0   & 0.4472   &      0  &       0\\
         0    &     0  &  0.4472   &      0\\
         0     &    0 &        0   & 0.4082\\
\end{bmatrix}.
\end{align}
From this one can have the stiffness matrix of the dual truss as:
\begin{align}
F'=\begin{bmatrix}
    0.7454 &        0   &      0 &        0\\
         0  &  0.6667  &       0  &       0\\
         0   &      0 &   0.6667   &      0\\
         0    &     0&         0    & 0.7454\\
\end{bmatrix}10^{-8}\frac{\text{m}}{\text{kN}}.
\end{align}

\begin{figure}[h]
\begin{center}
\includegraphics[width=80mm]{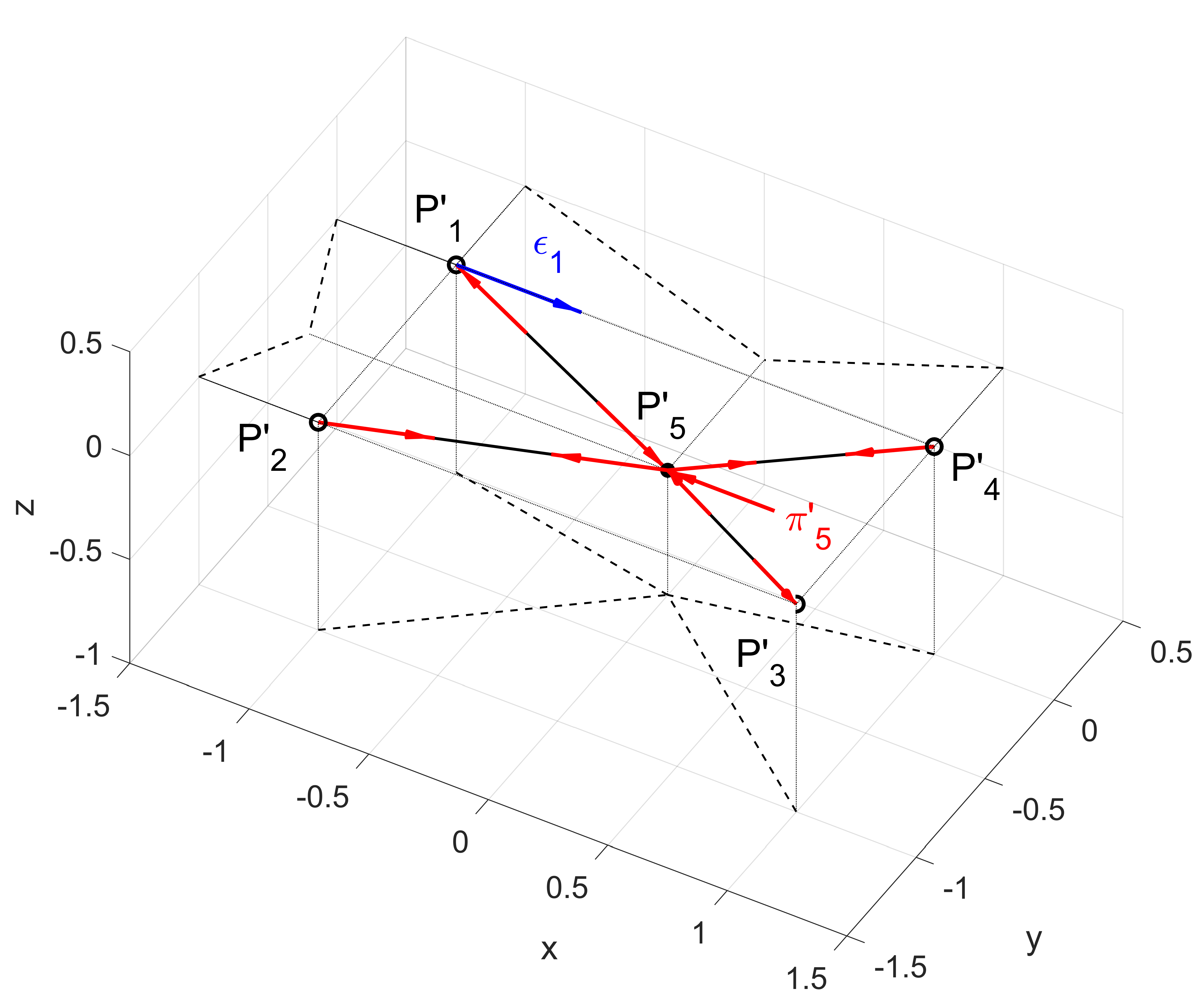}
\caption{Dual truss with 5 vertices and 4 internal bars. The supporting vertices are represented with $\circ$ while the internal one with $\bullet $.}\label{fig:racspelda}
\end{center}
\end{figure}

We will go with the strategy of $\vec{b}_{i,1}=(1,0,0)$, $\vec{b}_{i,2}=(0,1,0)$ and $\vec{b}_{i,3}=(0,0,1)$ for all $i$. Thus the relevant bases to this problem are $\{d_{1,j}\}$, $\{d_{5,j}\}$ and $\{f_{5,j}\}$. They are given by:

\begin{align}
 d_{1,1}&=(1, 0,0,0,0,0)\\
 d_{1,2}&=(0,0,0,0,0,-1)\\
d_{1,3}&=(0,0,0,0,1,0)\\
d_{5,1}&=(0,0,0,0,-0.4,0.2)\\
d_{5,2}&=(0,0.04,0.08,0.4,0,0)\\
d_{5,3}&=(0,0.08,0.16,-0.2,0,0)\\
f_{5,1}&=(0,-0.4,0.2,1,0,0)\\
f_{5,2}&=(0.4,0,0,0,1,0)\\
f_{5,3}&=(-0.2,0,0,0,0,1).
\end{align}
From this one can work out the relevant parts of $\Delta$ and $\Pi$. It turns out that $\Delta_1$ is the 3 dimensional identity matrix and
\begin{align}
 \Delta_5&=\begin{bmatrix}
    5 &        0   &      0\\
         0  &  1.1180   &      0\\
         0   &      0  &  4.4721\\
\end{bmatrix}\frac{1}{m}\\
\Pi_5&=\begin{bmatrix}
    0.4472 &        0 &        0\\
         0 &   0.4  &       0\\
         0 &        0  &  0.2\\
\end{bmatrix}.
\end{align}
The decomposition of the loads into the appropriate bases gives

\begin{align}
\delta_1=\begin{bmatrix}
   0.0017\\
   0\\
   0
\end{bmatrix} \text{rad} \ \text{and } \pi_5= \begin{bmatrix}
   -1\\
    0\\
    0
\end{bmatrix}\text{kN},
\end{align}
which can be transformed to truss loads as
\begin{align}
\pi'_5=\Pi_5^{-1}\pi_5=\begin{bmatrix}
-2.2361\\
   0\\
   0
\end{bmatrix} \text{kN}
\end{align}
and
\begin{align}
\epsilon_1=\Delta_1^{-1}\delta_1=\begin{bmatrix}
0.0017\\
   0\\
   0
\end{bmatrix} \text{m}.
\end{align}
Now everything is given in the problem of the dual truss, which can be seen in Figure  \ref{fig:racspelda}, along with the directions of the arising bar forces as well.

Solving the dual truss gives 
\begin{align}
\epsilon_5=\begin{bmatrix}
    0.0005\\
   -0.0009\\
   -0.0017
\end{bmatrix} \text{m} \ \text{and } \phi'= \begin{bmatrix}
   -4.5643\\
    5.5899\\
   -5.5899\\
    4.5643
\end{bmatrix}10^4 \ \text{kN},
\end{align}
which can be transformed back to the plate structure giving
\begin{align}
\delta_5=\begin{bmatrix}
    0.0025\\
   -0.0010\\
   -0.0078\\
\end{bmatrix} \text{and } \phi= \begin{bmatrix}
   -1.8634\\
    2.4999\\
   -2.4999\\
    1.8634
\end{bmatrix}10^4 \ \text{kN}.
\end{align}
Since the bases and dual bases are not normed, the actual rotation of $\m{P}_5$ is not $\norm{\delta_5}$, but one has to compute
\begin{align}
\begin{split}
d_5&=0.0175 d_{5,1}-0.0195 d_{5,2}-0.1561 d_{5,3}\\
&=(0,-0.0007,-0.0013,0.0012,-0.0010,0.0005).
\end{split}
\end{align}
The norm of this can be taken as defined in case of a line segment, returning the magnitude of the rotation of plate $5$ to be $0.0015$ radians or $0.0850$ degrees. The axis and direction of this is given by the line segment $d_5$ represents.
\section{Summary}
A metric correspondence was provided between forces and small displacements of a plate structure (made of plates rigid in their planes, soft perpendicular to their planes, joined by elastic edges) and those of its  dual: a truss with elastic bars. Since the computational methods of trusses are well developed and widely known, this ability to turn unfamiliar problems into familiar ones may give a useful tool in the hands of structural engineers or researchers. Especially now, when automation propels the spread of prefabricated plate structures, where the joining edges are usually softer then the plates itself.

\section*{Acknowledgments}
The author wishes to thank professors P.L.Várkonyi and T. Tarnai for useful discussions on the subject.



%
%
%
%
%
%
%
%
%


\bibliographystyle{default}
\bibliography{lemezirodalom}

\begin{thebibliography}{10}
\providecommand{\url}[1]{\texttt{#1}}
\providecommand{\urlprefix}{URL }
\expandafter\ifx\csname urlstyle\endcsname\relax
  \providecommand{\doi}[1]{DOI:\discretionary{}{}{}#1}\else
  \providecommand{\doi}{DOI:\discretionary{}{}{}\begingroup
  \urlstyle{rm}\Url}\fi
\providecommand{\eprint}[2][]{\url{#2}}

\bibitem{maxwell1870}
Maxwell JC.
\newblock I.—on reciprocal figures, frames, and diagrams of forces.
\newblock \emph{Transactions of the Royal Society of Edinburgh} 1870; 26(1):
  1–40.
\newblock \doi{10.1017/S0080456800026351}.

\bibitem{cremona1890graphical}
Cremona L and Beare T.
\newblock \emph{Graphical Statics: Two Treatises on the Graphical Calculus and
  Reciprocal Figures in Graphical Statics ...}
\newblock Claredon Press, 1890.

\bibitem{klein1904onMaxwell}
Klein F and Wieghardt K.
\newblock Über spannungsflächen und reziproke diagramme, mit besondere
  berücksichtigung der maxwellschen arbeiten.
\newblock In \emph{Gesammelte mathematische Abhandlungen}. Springer, 1922.
\newblock pp. 660--691.
\newblock
  \urlprefix\url{http://opac.sub.uni-goettingen.de/DB=1/PPN?PPN=237839962}.

\bibitem{crapo1982statics}
Crapo H and Whiteley W.
\newblock Statics of frameworks and motions of panel structures: a projective
  geometric introduction.
\newblock \emph{Structural Topology, 1982, n{\'u}m 6} 1982; .

\bibitem{wunderlich1982projective}
Wunderlich W.
\newblock Projective invariance of shaky structures.
\newblock \emph{Acta Mechanica} 1982; 42(3): 171--181.

\bibitem{TARNAI1989duality}
Tarnai T.
\newblock Duality between plane trusses and grillages.
\newblock \emph{International Journal of Solids and Structures} 1989; 25(12):
  1395 -- 1409.
\newblock \doi{https://doi.org/10.1016/0020-7683(89)90108-X}.

\bibitem{whiteley1991weavings}
Whiteley W.
\newblock Weavings, sections and projections of spherical polyhedra.
\newblock \emph{Discrete Applied Mathematics} 1991; 32(3): 275--294.

\bibitem{whiteley1989rigidity}
Whiteley W.
\newblock Rigidity and polarity.
\newblock \emph{Geometriae Dedicata} 1989; 30(3): 255--279.

\bibitem{whiteley1987rigidity}
Whiteley W.
\newblock Rigidity and polarity.
\newblock \emph{Geometriae Dedicata} 1987; 22(3): 329--362.

\bibitem{wester1989design}
\emph{Design of plate and lattice structures based on structural dualism},
  1989.

\bibitem{wester1990geodesic}
Wester T.
\newblock A geodesic dome-type based on pure plate action.
\newblock \emph{International Journal of Space Structures} 1990; 5(3-4):
  155--167.

\bibitem{bagger2010plate}
Bagger A, J{\"o}nsson J, Almegaard H et~al.
\newblock \emph{Plate shell structures of glass: Studies leading to guidelines
  for structural design}.
\newblock Technical University of Denmark (DTU), 2010.

\bibitem{Krieg2015robotic}
Krieg OD, Schwinn T, Menges A et~al.
\newblock Biomimetic lightweight timber plate shells: Computational integration
  of robotic fabrication, architectural geometry and structural design.
\newblock In Block P, Knippers J, Mitra NJ et~al. (eds.) \emph{Advances in
  Architectural Geometry 2014}. Cham: Springer International Publishing.
\newblock ISBN 978-3-319-11418-7, pp. 109--125.

\bibitem{BLANDINI2007addhesive}
Blandini L.
\newblock Structural use of adhesives for the construction of frameless glass
  shells.
\newblock \emph{International Journal of Adhesion and Adhesives} 2007; 27(6):
  499 -- 504.
\newblock \doi{https://doi.org/10.1016/j.ijadhadh.2006.09.001}.

\bibitem{SANTARSIERO2016laminated}
Santarsiero M, Louter C and Nussbaumer A.
\newblock Laminated connections for structural glass applications under shear
  loading at different temperatures and strain rates.
\newblock \emph{Construction and Building Materials} 2016; 128: 214 -- 237.
\newblock \doi{https://doi.org/10.1016/j.conbuildmat.2016.10.045}.

\bibitem{ZANGENBERG201268}
Zangenberg J, Poulsen S, Bagger A et~al.
\newblock Embedded adhesive connection for laminated glass plates.
\newblock \emph{International Journal of Adhesion and Adhesives} 2012; 34: 68
  -- 79.
\newblock \doi{https://doi.org/10.1016/j.ijadhadh.2012.01.003}.

\bibitem{li2015segmental}
Li JM and Knippers J.
\newblock Segmental timber plate shell for the landesgartenschau exhibition
  hall in schwäbisch gmünd—the application of finger joints in plate
  structures.
\newblock \emph{International Journal of Space Structures} 2015; 30(2):
  123--139.
\newblock \doi{10.1260/0266-3511.30.2.123}.

\bibitem{stitic2018experimental}
Stitic A, Robeller C and Weinand Y.
\newblock Experimental investigation of the influence of integral mechanical
  attachments on structural behaviour of timber folded surface structures.
\newblock \emph{Thin-Walled Structures} 2018; 122: 314 -- 328.
\newblock \doi{https://doi.org/10.1016/j.tws.2017.10.001}.

\bibitem{konstantatou2016reciprocal}
\emph{Reciprocal constructions using Ponclelet duality and conic section},
  2016.

\bibitem{mcrobie2016maxwell}
McRobie A.
\newblock Maxwell and rankine reciprocal diagrams via minkowski sums for
  two-dimensional and three-dimensional trusses under load.
\newblock \emph{International Journal of Space Structures} 2016; 31(2-4):
  203--216.

\bibitem{Mitchell2016partI}
Mitchell T, Baker W, McRobie A et~al.
\newblock Mechanisms and states of self-stress of planar trusses using graphic
  statics, part i: The fundamental theorem of linear algebra and the airy
  stress function.
\newblock \emph{International Journal of Space Structures} 2016; 31(2-4):
  85--101.
\newblock \doi{10.1177/0266351116660790}.

\bibitem{McRobie2016partII}
McRobie A, Baker W, Mitchell T et~al.
\newblock Mechanisms and states of self-stress of planar trusses using graphic
  statics, part ii: Applications and extensions.
\newblock \emph{International Journal of Space Structures} 2016; 31(2-4):
  102--111.
\newblock \doi{10.1177/0266351116660791}.

\bibitem{McRobie2017elastographics}
McRobie A, Konstantatou M, Athanasopoulos G et~al.
\newblock Graphic kinematics, visual virtual work and elastographics.
\newblock \emph{Royal Society Open Science} 2017; 4(5).
\newblock \doi{10.1098/rsos.170202}.

\bibitem{lamagna2012nature}
La~Magna R, Waimer F and Knippers J.
\newblock Nature-inspired generation scheme for shell structures.
\newblock In \emph{Proceedings of the International Symposium of the IASS-APCS
  Symposium, Seoul, South Korea}.
\newblock \urlprefix\url{http://dx.doi.org/10.18419/opus-105}.

\bibitem{Livesley}
Livesley R.
\newblock \emph{Matrix Methods of Structural Analysis (Second Edition)}.
\newblock Second edition ed. Pergamon International Library of Science,
  Technology, Engineering and Social Studies, Pergamon, 1975.
\newblock ISBN 978-0-08-018888-1.
\newblock \doi{https://doi.org/10.1016/B978-0-08-018888-1.50001-0}.

\bibitem{roller_szabo}
Szabó J and Roller B.
\newblock \emph{Anwendung der Matrizenrechnung auf Stabwerke}.
\newblock Budapest: Akadémiai Kiadó, 1978.
\newblock ISBN 963-05-1216-5.

\bibitem{pottmann2001computational}
Pottmann H and Wallner J.
\newblock \emph{Computational Line Geometry}.
\newblock Mathematics and Visualization, Springer Berlin Heidelberg, 2001.
\newblock ISBN 9783540420583.

\bibitem{clebsch1906mathematische}
Clebsch A, Neumann C, Klein F et~al.
\newblock \emph{Mathematische Annalen}.
\newblock 62. k., J. Springer, 1906.

\bibitem{ball1998treatise}
Ball R.
\newblock \emph{A Treatise on the Theory of Screws}.
\newblock Cambridge Mathematical Library, Cambridge University Press, 1998.
\newblock ISBN 9780521636506.

\bibitem{davidson2004robots}
Davidson J and Hunt K.
\newblock \emph{Robots and Screw Theory: Applications of Kinematics and Statics
  to Robotics}.
\newblock Oxford University Press, 2004.
\newblock ISBN 9780198562450.

\bibitem{gallardo2016kinematic}
Gallardo-Alvarado J.
\newblock \emph{Kinematic Analysis of Parallel Manipulators by Algebraic Screw
  Theory}.
\newblock Springer International Publishing, 2016.
\newblock ISBN 9783319311265.

\end{thebibliography}

\end{document}